\documentclass[11pt,a4paper]{article}
\usepackage{jheppub}
\usepackage[T1]{fontenc}
\usepackage{amssymb}
\usepackage{mathtools}

\usepackage{stackengine}

\usepackage{scalerel}

\usepackage{hyperref}

\usepackage{epsfig}
\usepackage{amsmath,latexsym,amssymb}
\usepackage{graphicx}
\usepackage[latin1]{inputenc}
\usepackage{braket}
\usepackage{pdfsync}

\usepackage{framed}

\usepackage{mathtools}

\def\be{\begin{equation}}
\def\ee{\end{equation}}
\def\ba{\begin{eqnarray}}
\def\ea{\end{eqnarray}}

\newcommand{\bz}{\bar{z}}

\newcommand{\bh}{\bar{h}}

\newcommand{\bc}{\bar{c}}
\newcommand{\bd}{\bar{d}}

\newcommand{\zbar}{\bar{z}}

\def\d{\delta}
\def\D{\Delta}
\def\e{\epsilon}

\def\m{\mu}

\def\om{\omega}

\def\l{\lambda}
\def\L{\Lambda}

\def\cM{{\cal M}}

\def\cO{{\cal O}}

\def\cA{{\cal A}}

\newcommand{\comment}[1]{}

\newcommand{\eea}{\end{eqnarray}}
\def\lf{\left}
\def\ri{\right}

\def\Tr{{\rm Tr}}

\author{Wei Fan${}^{1}$,
Angelos Fotopoulos${}^{1,2}$ and
Tomasz R.\ Taylor${}^{1}$ \\[0.5cm]
 $^1${\it Department of Physics \\
  Northeastern University, Boston, MA 02115, USA}\\[0.2cm]
 $^2${\it Department of Natural Sciences\\  Assumption College, Worcester, MA 01609, USA}
 }

\title{\boldmath Soft Limits of Yang-Mills Amplitudes\\[2mm] and Conformal Correlators  \unboldmath}

\abstract{We study tree-level celestial amplitudes in Yang-Mills theory --  Mellin transforms of multi-gluon scattering amplitudes that convert them into the correlators of conformal primary fields on two-dimensional celestial sphere. By using purely field-theoretical methods, we show that the soft conformal limit of celestial amplitudes, in which one of the primary field operators associated to gauge bosons becomes a dimension one current, is dominated by the contributions of low-energy soft particles. This result confirms conclusions reached by using Yang-Mills theory formulated in curvilinear coordinates, as pioneered by Strominger. By using well-known collinear limits of Yang-Mills amplitudes, we derive the OPE rules for the primary fields and the holomorphic currents arising in the conformally soft limit. The Ward identities following from OPE have the same form as the identities derived by using soft theorems.}


\bigskip

\begin{document}
\maketitle

\section{Introduction}\label{intro}
The scattering processes of elementary particles are usually described in terms of amplitudes that determine transition probabilities depending on kinematic variables and internal charges. One of their fundamental properties is Lorentz invariance under $SO(1,3)\sim SL(2,\mathbb{C})$ symmetry group\footnote{More precisely, $SO^+(1,3)\sim PSL(2,\mathbb{C})$.}  which, together with unitarity, imposes very strong constraints on their kinematic dependence. $SL(2,\mathbb{C})$ is also a symmetry group of conformal geometry on the Riemann sphere that can be identified, by using Bondi coordinates, as the celestial sphere at null infinity. Hence it is not surprising that the scattering amplitudes can be recast into ``celestial'' amplitudes having the form of conformal correlation functions on the celestial sphere.

There are two roads leading to celestial amplitudes. One, pioneered by Strominger \cite{Strominger:2017zoo}, is by formulating quantum field theory in curvilinear (Bondi) coordinates which are best suited for investigating the asymptotic structure of spacetime. This may well be an important step towards developing a holographic description of flat spacetime. Another one is by starting from amplitudes describing transitions between momentum eigenstates and changing the asymptotic basis from plane waves to the so-called conformal wave-packets with well-defined conformal weights \cite{Pasterski:2016qvg,Pasterski:2017kqt,Pasterski:2017ylz,
Schreiber:2017jsr,Stieberger:2018edy}. This is accomplished by taking Mellin transforms of traditional amplitudes and utilizes standard tools of four-dimensional quantum field theory and two-dimensional ``celestial'' CFT, without referring to curvilinear coordinates. This road can be followed \cite{Stieberger:2018onx} for investigating connections between gauge theories and gravity, aiming towards an explanation of the ``double copy''  or ``Einstein=Yang-Mills$^2$'' relations \cite{Bern:2008qj,Bern:2010ue}. The first road led to a startling discovery that the soft theorems describing emissions of low-energy photons and gluons take the form of Ward identities  associated to two-dimensional CFT currents \cite{Strominger:2017zoo,Lysov:2014csa,Cheung:2016iub}. These currents can be identified
as dimension one limits of primary fields associated to gauge bosons, as they appear in the so-called {\em conformally\/} soft limit \cite{Donnay:2018neh} of the amplitudes. In this paper we follow the second, old-fashioned field-theoretical route to show that the {\em conformally\/} soft limit of tree-level celestial Yang-Mills amplitudes is dominated by the contributions of low-energy, soft particles.
We study the corresponding CFT correlators and derive the OPE rules for the primary fields from well-known collinear limits of invariant matrix elements. We show that OPE yields the same Ward identities  as the identities derived by using soft theorems \cite{Cheung:2016iub}.

The paper is organized as follows. In section \ref{Prelim} we give a short account of the notation and basic formulas describing conformal primary wavefunctions and their properties. In section \ref{MellinAmplitudes}, we
discuss soft limits of celestial amplitudes obtained by Mellin transformations converting the plane wave basis into conformal wave-packets.
Usually, the conformally soft limit is related in a straightforward way to the low-energy soft limit however, in some cases, especially with a smaller number of particles ($n<6$) this limit is more subtle. It requires either using a different basis for the solutions of kinematic constraints or exploring the ``corners'' of kinematic space that are accessible only through some special solutions of kinematic constraints. In all cases, the soft and conformally soft limits are equivalent.
 In section \ref{sec:OPE}, we derive the OPE rules for the primary conformal field operators associated to gauge bosons. They follow from the well-known collinear limits of invariant matrix elements \cite{tt}. We then take the limit when one of the operators in the product becomes conformally soft (dimension one) and obtain the OPE rules for the holomorphic and anti-holomorphic currents. The Ward identities following from such OPE have exactly the same form as those obtained by using soft theorems. In the Appendix, we give a solution of $n$-particle kinematic constraints best suited for investigating the soft limits.

 The present paper has some overlap with Ref.\cite{prs} which
 reaches similar conclusions regarding soft and conformally soft limits.

\section{Preliminaries}\label{Prelim}
In this section we establish notation and give a short exposition of the properties of conformal primary wavefunctions as discussed in Refs.\cite{Pasterski:2017kqt,Pasterski:2017ylz}.

In four-dimensional Minkowski space, the Lorentz group is equivalent to the two-dimensional conformal group $SL(2,\mathbb{C})$. In order to elucidate this connection, it is convenient to use Bondi coordinates $(u,r,z, \bz)$, where the coordinates $z,\bz$ parametrize what is known as the celestial sphere ${\cal C S}^2$ at null infinity. On ${\cal C S}^2$, the $SL(2,\mathbb{C})$ Lorentz group acts as the global conformal symmetry group:
\be\label{SLz}
z\to {a z+b\over c z+d}, \qquad ad-bc=1.
\ee

A general light-like momentum vector can be parametrized as
\be\label{momparam}
p^\mu= \omega q^\mu, \qquad q^\mu={1\over 2} (1+|z|^2, z+\bz, -i(z-\bz), 1-|z|^2)
\ee
where $q^\mu$ is a null vector, the direction along which the massless state propagates, and $\omega$ is the light cone energy. Their transformation properties under the  Lorentz group are
\be\label{SLq}
\omega \to (c z+d) (\bc \bz +\bd) \omega, \quad q^\mu \to q'^\mu=  (c z+d)^{-1} (\bc \bz +\bd)^{-1}  \L^\mu_{\ \nu} q^\nu
\ee
so that $p^\mu \to p'^\mu=\L^\mu_{\ \nu}  p^\nu$, where the matrix $\L^\mu_{\ \nu} $ is the associated Lorentz group element in the four-dimensional representation. The usual gauge boson plane waves, in the Lorentz gauge $\partial_\mu A^\mu=0$, are
\be\label{planewave}
\e_{\mu \ell}(p) e^{\mp i |p_0| X^0  \pm i \vec{p} \cdot \vec{X}}, \qquad \ell=\pm 1
\ee
where $\e_{\mu \ell}(p)$  is the polarization  vector, $\ell$ is the helicity and the $\pm$ sign in the exponential is used to distinguish between incoming and outgoing solutions.

As explained in \cite{Pasterski:2017kqt,Pasterski:2017ylz}, a Mellin transform allows us to construct another basis of massless gauge boson solutions
\be\label{confprim}
A^{\D  \pm}_{\mu J} (X^\mu, z, \bz)
\ee
labelled by the points $z,\bz$ on ${\cal C S}^2$, the conformal dimension $\D$ and two-dimensional spin $J=\pm 1$, with  the $\pm$ superscript used to distinguish between outgoing and incoming wavefunctions. The conformal spin can be identified with four-dimensional helicity.
These solutions satisfy Maxwell equations and transform under the Lorentz group $SL(2,\mathbb{C})$  as four dimensional vectors and two dimensional conformal (quasi) primaries of spin $J=\pm 1$ and weight $\D$:
\be\label{confprimtrans}
A^{\D  \pm}_{\mu J} \Big(\L_\mu^{\ \nu} X^\nu, {a z + b\over c z +d}, {\bar{a} \bz +\bar{b} \over \bc \bz +\bd}\Big)= (c z +d)^{\D +J}(\bc \bz +\bd)^{\D -J}\ \L_\mu^{\ \nu}A^{\D  \pm}_{\nu J}(X^\mu, z,\bz)
\ee
We will be often characterizing such fields by their conformal weights
\be h= {1\over 2} (\D+J), \quad \bh=  {1\over 2} (\D-J)\ .\ee

The polarization vectors $\e_{\mu l}(p)$ of the one-particle massless states can be written in terms of the null vector $q^\mu$
\be\label{polar}
\partial_z q^\mu =  \e^\mu _+(p) \qquad \partial_{\bz}  q^\mu =  \e^\mu _-(p)
\ee
The conformal primary wavefunctions are given by\footnote{ The $+i \e$ prescription is used to circumvent the $q\cdot X$ singularity. The same $\e$ acts as an ultraviolet regulator in Mellin transforms discussed below.}
\be\label{confprimexpr1}
A^{\D  \pm}_{\mu J} (X^\mu, z, \bz)=  {\partial_J q^\mu \over (- q\cdot X \mp i \e)^\D}+   {\partial_J q\cdot X \over (- q\cdot X \mp i \e)^{\D+1}} q^\mu\ .
\ee
They satisfies both the Lorentz  and radial gauge conditions
\be\label{eq:gaugecond}
\partial_\m A^{\D  \pm}_{\mu J}=0 \ , \qquad  X^\m A^{\D  \pm}_{\mu J}=0
\ee
and have dimensions $\D = 1+i \l $ with $\l \in \mathbb{R}$. In Eq.(\ref{confprimexpr1}), $\partial_J$ denotes $\partial_z  / \partial_{\bz} $ for  positive ($J=+1$) / negative ($J=-1$) helicity respectively. These states belong to the principal continuous series  of the unitary representations of $SL(2, \mathbb{C})$.

The connection to Mellin-transformed plane waves is established in the following way. The Mellin transform of the plane wave (\ref{planewave}) is
\be\label{confprimMellin}
V^{\D  \pm}_{\mu J} (X^\mu, z, \bz)\equiv \partial_J q_\mu \int _0^\infty d\omega  \ \omega^{\D-1} e^{\mp i \omega q \cdot  X -\e \omega\ .}
\ee
The conformal wave function (\ref{confprimexpr1}) can be written as
\be\label{confprimexpr3}
A^{\D  \pm}_{\m J}= g(\D){V^{\D  \pm}_{\mu J} }+ \partial_\m a_J^{\D \pm}
\ee
where
\be\label{gdef}g(\D)=(\pm i)^\D{\D -1 \over  \Gamma(\D+1) }=(\pm i)^{1+i\l}{i\lambda\over\Gamma(2+i\lambda)}\ee
and
\be\label{agauge}
a_J^{\D\pm}(X^\mu, z, \bz)= \left(  {\partial_J q\cdot X \over \D(- q\cdot X \mp i \e)^{\D}} \right).
\ee
This means that the conformal wave function is equivalent to a Mellin-transformed plane wave up to an additive gauge transformation, and up to a multiplicative normalization factor $g(\D)$ that will play important role in the following discussion. In order to simplify the formulas, whenever using $g(\D)$ of Eq.(\ref{gdef}), we will use the $(- i)^\D$ factor corresponding to outgoing (+) wavefunctions. At the end of section 4 we will explain the role of this factor in implementing the CPT symmetry of four-dimensional theory.

In this paper, we will analyze the soft {\em conformal\/} limit of $\lambda\to 0 ~~(\D\to 1)$. This limit is singular, for two reasons. First of all, the normalization factor $g\to 0$. Furthermore, for $\D=1$, the gauge parameter $a_J^{\D\pm}$ of Eq.(\ref{agauge}) is {\em not\/} suppressed at $X\to\infty$, therefore it corresponds to a {\em large\/} gauge transformation that can have observable effects if the gauge sources do not vanish at the space-time boundary. From now on, this mode will be called the ``Goldstone mode.'' One of our goals is to elaborate on the connection between the {\em conformal\/} soft limit and the $\omega\to 0$ zero energy limit described by the well-known soft theorems.

\section{Soft vs Conformally Soft Limits of Celestial Yang-Mills Amplitudes}\label{MellinAmplitudes}
Since according to Eq.(\ref{confprimexpr3}), the conformal primary wave functions (\ref{confprimexpr1}) correspond, up to pure gauge terms,  to Mellin transforms of plane wave packets, Yang-Mills amplitudes can be transformed into  ``celestial'' basis by applying Mellin transformations with respect to the energy variables:
\be\label{MellinAmplnw}
 {\cA}_{J_1\dots J_n}(\D_i, z_i, \bz_i)=\left(  \prod_{i=1}^n g(\lambda_i)\int d \omega_i \ \omega_i^{i \l_i} \right) \d^{(4)}(\sum_i \e_i \om_i q_i)
 \cM_{\ell_1\dots \ell_n}(\omega_i, z_i, \bz_i)
\ee
where $\e_i=\pm$ for outgoing/ingoing particles. Here, $\cM$ is the Feynman invariant matrix element for the scattering process. We focus on ``partial'' amplitudes associated to one particular Chan-Paton group factor $\Tr(T^1T^2\cdots T^n)$. For a review of Yang-Mills amplitudes, see Ref\cite{tt}. The conformal dimensions $\D_i=1+i\lambda_i$ are Mellin-dual to the energies $\omega_i$.
Under $SL(2,\mathbb{C})$ transformations, celestial amplitudes transform as two-dimensional conformal correlators:
\be\label{AmplTrans}
{\cA}_{J_1\dots J_n}\Big(\D_i, {a z_i + b\over c z_i +d}, {\bar{a} \bz_i +\bar{b} \over \bc \bz_i +\bd}\Big)= \prod_{i=1}^n \ ( c z_i +d)^{\D_i+J_i} (\bc \bz_i +\bd)^{\D_i-J_i} {\cA}_{J_1\dots J_n}(\D_i, z_i, \bz_i)\ ,
\ee
In this way, four-dimensional scattering amplitudes are recast into the form of correlation functions of primary CFT field operators  on ${\cal C S}^2$.

Until quite recently, most of attention was focused on the correlators involving one or more operators associated to ``soft'' gauge bosons with Minkowski energies $\omega\to 0$. In Ref.\cite{Cheung:2016iub}, it was argued that this limit corresponds to the {\em conformally\/} soft limit of $\lambda\to 0$ ($\D\to 1$) of the dual variable. Then the soft operators can be identified as $\D=1$ CFT currents and the well-known soft theorems appear as Ward identities associated to such currents. Beyond the soft limit,  several examples of Yang-Mills amplitudes have been recently discussed in Refs.\cite{Pasterski:2017ylz,Schreiber:2017jsr,Stieberger:2018edy}. In this section, we discuss the conformally soft limit of such ``hard'' amplitudes.

In Eq.(\ref{MellinAmplnw}), the invariant matrix element $\cM$ is expressed in terms of the energy $\omega$ and ``angular'' $z$ variables. It is very instructive to discuss the number of independent ``energy'' and ``angle'' variables. An $n$-point scattering amplitudes depends on $3n{-}10$ Lorentz-invariant kinematic variables. For $n=4$, we have 2 variables that can be expressed in terms of one (light-cone) energy scale and one angle, specified by (the real part of) one complex coordinate on ${\cal C S}^2$, say $z_4$,  or equivalently by (the real part of) one cross-ratio. For $n=5$, there is still only one energy scale but two complex points $z_4, z_5$. For $n=6$, we have three complex points, therefore we have two energy scales, for the total of 8 variables. This is the lowest $n$ case when one can discuss the soft limit in a straighforward way, by taking a direct low energy limit, e.g.\ $\omega_6\to 0$, without ``engineering'' it by some special configurations of celestial points.

In order to make the above arguments more explicit, we discuss the momentum conservation constraints implied by the delta function inside Eq.(\ref{MellinAmplnw}). For $n=6$, it is convenient to use the following set of cross ratios:
\begin{equation}
    t_4 = \frac{z_{12}z_{34}}{z_{13}z_{24}}, ~~ t_5=\frac{z_{12}z_{35}}{z_{13}z_{25}}, ~~ t_6=\frac{z_{12}z_{36}}{z_{13}z_{26}}\ ,
\end{equation}
where $z_{ij}\equiv z_i-z_j$.
A straightforward but tedious computation allows rewriting the delta function as
\begin{equation}\label{delta6}
    \delta^4\big(\sum_{i=1}^{6} \epsilon_i \omega_i q_i\big) = \frac{i}{4}\frac{(1-t_4)(1-\bar{t}_4)}{t_4-\bar{t}_4} \frac{1}{|z_{14}|^2|z_{23}|^2} \prod_{i=1}^{4}\delta(\omega_i-\omega_i^\star),
\end{equation}
where the solutions $\omega_i^\star$ are
\begin{align}
    \omega_1^\star&=t_4 \Big|\frac{z_{24}}{z_{12}}\Big|^2 \frac{(1-t_4)(1-\bar{t}_4)}{t_4-\bar{t}_4}\left(\epsilon_1\epsilon_5\frac{t_5-\bar{t}_5}{(1-t_5)(1-\bar{t}_5)}\Big|\frac{z_{15}}{z_{14}}\Big|^2\omega_5 + \epsilon_1\epsilon_6 \frac{t_6-\bar{t}_6}{(1-t_6)(1-\bar{t}_6)}\Big|\frac{z_{16}}{z_{14}}\Big|^2\omega_6\right) \nonumber\\
    &{\quad } - \epsilon_1\epsilon_5 t_5\Big|\frac{z_{25}}{z_{12}}\Big|^2\omega_5 - \epsilon_1\epsilon_6 t_6\Big|\frac{z_{26}}{z_{12}}\Big|^2\omega_6 \\
    &= f_{15} \omega_5 + f_{16} \omega_6\nonumber \\
    \omega_2^\star&=- \frac{1-t_4}{t_4} \Big|\frac{z_{34}}{z_{23}}\Big|^2 \frac{(1-t_4)(1-\bar{t}_4)}{t_4-\bar{t}_4}\left(\frac{\epsilon_1\epsilon_5}{\epsilon_1\epsilon_2}
    \frac{t_5-\bar{t}_5}{(1-t_5)(1-\bar{t}_5)}\Big|\frac{z_{15}}{z_{14}}\Big|^2\omega_5 + \frac{\epsilon_1\epsilon_6}{\epsilon_1\epsilon_2} \frac{t_6-\bar{t}_6}{(1-t_6)(1-\bar{t}_6)}\Big|\frac{z_{16}}{z_{14}}\Big|^2\omega_6\right)\nonumber \\
    &{\quad } + \frac{\epsilon_1\epsilon_5}{\epsilon_1\epsilon_2} \frac{1-t_5}{t_5}\Big|\frac{z_{35}}{z_{23}}\Big|^2\omega_5 + \frac{\epsilon_1\epsilon_6}{\epsilon_1\epsilon_2} \frac{1-t_6}{t_6}\Big|\frac{z_{36}}{z_{23}}\Big|^2\omega_6 \\
    &= f_{25} \omega_5 + f_{26} \omega_6 \nonumber\\
    \omega_3^\star&= (1-t_4)\Big|\frac{z_{24}}{z_{23}}\Big|^2 \frac{(1-t_4)(1-\bar{t}_4)}{t_4-\bar{t}_4}\left(\frac{\epsilon_1\epsilon_5}{\epsilon_1\epsilon_3}\frac{t_5-\bar{t}_5}{(1-t_5)(1-\bar{t}_5)}\Big|\frac{z_{15}}{z_{14}}\Big|^2\omega_5 + \frac{\epsilon_1\epsilon_6}{\epsilon_1\epsilon_3} \frac{t_6-\bar{t}_6}{(1-t_6)(1-\bar{t}_6)}\Big|\frac{z_{16}}{z_{14}}\Big|^2\omega_6\right)\nonumber \\
    &{\quad } - \frac{\epsilon_1\epsilon_5}{\epsilon_1\epsilon_3} (1-t_5)\Big|\frac{z_{25}}{z_{23}}\Big|^2\omega_5 - \frac{\epsilon_1\epsilon_6}{\epsilon_1\epsilon_3} (1-t_6)\Big|\frac{z_{26}}{z_{23}}\Big|^2\omega_6 \\
    &= f_{35} \omega_5 + f_{36} \omega_6\nonumber \\
    \omega_4^\star&= - \frac{(1-t_4)(1-\bar{t}_4)}{t_4-\bar{t}_4}\left(\frac{\epsilon_1\epsilon_5}{\epsilon_1\epsilon_4}
    \frac{t_5-\bar{t}_5}{(1-t_5)(1-\bar{t}_5)}\Big|\frac{z_{15}}{z_{14}}\Big|^2\omega_5 + \frac{\epsilon_1\epsilon_6}{\epsilon_1\epsilon_4} \frac{t_6-\bar{t}_6}{(1-t_6)(1-\bar{t}_6)}\Big|\frac{z_{16}}{z_{14}}\Big|^2\omega_6\right)\nonumber \\
    &= f_{45} \omega_5 + f_{46} \omega_6
\end{align}
Note that the coefficients $f_{kj}, ~k=1,2,3,4$ depend on $z_j$ and $z_1,z_2,z_3,z_4$ only. In the Appendix, we give similar expressions for arbitrary $n\ge 5$. They involve $n{-}4$ independent energy variables and $n{-}3$ complex cross ratios, adding up to $3n{-}10$ kinematic variables.

In order to discuss the soft limit, we focus on the $n=6$ ``mostly plus'' MHV amplitude\footnote{Helicity amplitudes are expressed in terms of $(\omega,z)$ variables by using $\langle i\, j\rangle=\sqrt{\omega_{i}\omega_j} z_{ij}$, where $z_{ij}=z_i-z_j$.}
\be\label{mhv6}\cM_{--++++}(\omega_i, z_i)={\omega_1\omega_2\over\omega_3\omega_4\omega_5\omega_6}\frac{z_{12}^3}{z_{23}
z_{34}z_{45}z_{56}z_{61}} \ee
After inserting it into the Mellin integral (\ref{MellinAmplnw}) and using Eq.(\ref{delta6}), we obtain
\begin{align}
    \int_0^\infty &\big(\prod_{i=1}^{6} d\omega_i \omega_i^{i\lambda_i}\big) \frac{\omega_1\omega_2}{\omega_3\omega_4\omega_5\omega_6} \frac{z_{12}^3}{z_{23}z_{34}z_{45}z_{56}z_{61}} \frac{i}{4}\frac{(1-t_4)(1-\bar{t}_4)}{t_4-\bar{t}_4} \frac{1}{|z_{14}|^2|z_{23}|^2} \prod_{i=1}^{4}\delta(\omega_i-\omega_i^\star)\nonumber\\
    &=\frac{i}{4}\frac{(1-t_4)(1-\bar{t}_4)}{t_4-\bar{t}_4} \frac{1}{|z_{14}|^2|z_{23}|^2} \frac{z_{12}^3}{z_{23}z_{34}z_{45}z_{56}z_{61}}\, \mathcal{I}_6
\end{align}
where
\begin{align}
    \mathcal{I}_6&=\int_0^\infty d\omega_5 d\omega_6\, \omega_5^{-1+i\lambda_5} \omega_6^{-1+i\lambda_6} (f_{15}\omega_5+f_{16}\omega_6)^{1+i\lambda_1} (f_{25}\omega_5+f_{26}\omega_6)^{1+i\lambda_2}\nonumber\\
    &{\qquad } \times (f_{35}\omega_5+f_{36}\omega_6)^{-1+i\lambda_3} (f_{45}\omega_5+f_{46}\omega_6)^{-1+i\lambda_4}.\label{int6}
\end{align}
It is clear that the above integral is singular in the soft conformal limits $\lambda_5\to 0$ and $\lambda_6\to 0$. It contains single poles that can be exhibited by changing the integration variables to
\begin{equation}\label{change6}
    \omega_t=\omega_5+\omega_6 \Rightarrow \int_0^\infty d\omega_5 d\omega_6\cdots = \int_0^\infty d\omega_t \int_0^{\omega_t} d\omega_6\cdots .
\end{equation}
Then
\begin{align}
    \mathcal{I}_6&=\int_0^\infty d\omega_t \int_0^{\omega_t} d\omega_6\,  \omega_6^{-1+i\lambda_6}(\omega_t-\omega_6)^{-1+i\lambda_5}  (f_{15}\omega_t+(f_{16}-f_{15})\omega_6)^{1+i\lambda_1} (f_{25}\omega_t+(f_{26}-f_{25})\omega_6)^{1+i\lambda_2}\nonumber\\
    &{\qquad } \times (f_{35}\omega_t+(f_{36}-f_{35})\omega_6)^{-1+i\lambda_3} (f_{45}\omega_t+(f_{46}-f_{45})\omega_6)^{-1+i\lambda_4}.\label{int6f}
\end{align}
The leading $1/\lambda_6$ pole originates from the (Minkowski) soft region of $\omega_6\approx 0$:
\begin{align}\mathcal{I}_6(\lambda_6\to 0)\longrightarrow &
\int_0^\infty d\omega_t \int_0^{\omega_t} d\omega_6\, \omega_6^{-1+i\lambda_6}
{\omega_t}^{-1+i\lambda_5}
(f_{15}\omega_t)^{1+i\lambda_1} (f_{25}\omega_t)^{1+i\lambda_2} (f_{35}\omega_t)^{-1+i\lambda_3} (f_{45}\omega_t)^{-1+i\lambda_4}\nonumber \\
& = \frac{1}{i\lambda_6}\,\mathcal{I}_5 + \makebox{finite}.\label{softcc6}
\end{align}
Similarly,  the leading $1/\lambda_5$ pole originate from the upper limit $\omega_6\approx\omega_t$ which corresponds to $\omega_5\to 0$.

 The above discussion makes it very clear that the {\em conformal soft singularities are due to soft particles with (almost) vanishing energies\/}. Taking into acount that the normalization factors $g(\lambda)\to \lambda$, c.f.\ Eqs.(\ref{MellinAmplnw}), (\ref{gdef}) and
\be\label{mhv5}\cM_{--+++}(\omega_i, z_i)={\omega_1\omega_2\over\omega_3\omega_4\omega_5}\frac{z_{12}^3}{z_{23}
z_{34}z_{45}z_{51}} \ ,\ee
  one finds that the celestial amplitudes have well-defined conformally soft limits, which in this case is
\be\cA_{--+++(+)}=(-i)\Big(\frac{1}{z_{56}}+\frac{1}{z_{61}}\Big)\cA_{--+++}\ee
where the helicity index in parentheses denotes the limit $\Delta=1$ of the corresponding conformal dimension.

Looking back at the  case of $n=6$, one may ask the question how to extract the conformally soft singularities for the remaining particles, $\lambda_i\to 0$ with $ i<5$. The obvious answer is that one has to solve the delta function constraints (\ref{delta6}) with a basis of four energy variables different from $\omega_i^*, ~i=1,2,3,4$. In this way, one can identify all $\lambda\to 0$ singularities of Mellin integrals, in each case associating it to the soft limit of one of energies from outside the basis set. This means that the solutions like (\ref{delta6}) miss the cases when one of the basis energies becomes zero.

It is straightforward to extend the above discussion to $n>6$ by using the solutions of kinematic constraint written in the Appendix.  The limit of $\lambda\to 0$  is always associated to soft gauge bosons. Before discussing this limit in the context of two-dimensional CFT, let us comment on the special case of $n=4$ which was discussed extensively in Refs.\cite{Pasterski:2017ylz,Stieberger:2018edy}. After making connection between soft and conformally soft limits, we can extract the soft conformal limits by localizing the Mellin integrals on the low energy $\omega\to 0$ regions of the corresponding particles. This case, similarly to four basis energies encountered in the case of $n=6$, requires though a more careful handling of the kinematic constraints.
We will show that, as expected, the $n=4$ amplitude reduces in the conformally soft limit to a three-point amplitude.

{}In the case of three external particles, the amplitudes vanish due to  kinematic constraints ($3n{-}10{=}-1$). These constraints can be relaxed by changing the metric signature from $(+--\, -)$ to $(+-+\, -)$. This allows treating $z$ and $\zbar$ as two {\it independent\/} real variables. Then two classes of non-trivial kinematic solutions are allowed: all $z_{ij}=0$ with all $\zbar_{ij}\neq 0$ or all $\zbar_{ij}=0$ with all $z_{ij}\neq 0$. In the case of  amplitudes involving three gauge bosons, the first one is appropriate for  ``mostly minus'' helicity configurations while the second one is good for the ``mostly plus'' amplitudes. We will focus on the latter ones. Assuming all $z_{ij}\neq 0$, and specifying to the case of $\epsilon_1=\epsilon_2=-\epsilon_3=1$, the momentum-conserving delta function can be written as
\be\label{deltathree}\delta^{(4)}(\omega_1 q_1+\omega_2 q_2-\omega_3 q_3) ={4\over \omega_3^2}\ {1\over z_{23}z_{31}}\ \delta(\omega_1-{z_{32}\over z_{12}}\omega_3)\ \delta(\omega_2-{z_{31}\over z_{21}}\omega_3)\ \delta(\zbar_{13})\ \delta(\zbar_{23})\ ,\ee
with the additional constraint
that the variables must be ordered in one of two possible ways:
$z_1< z_3<z_2$ or $z_2< z_3<z_1$, in order to ensure that all energies are positive.
Using the well-known three-particle MHV amplitude, one obtains
\be\label{MHVcorrthree}
 {\cA}_{--+}= \left( 4 \prod_{i=1}^3 g(\lambda_i)\right) z_{21}^{1-i(\l_1+\l_2)} z_{23}^{i\l_1-1} z_{31}^{i\l_2-1}\d(\bz_{13})\d(\bz_{23}) \int d\om_3 \ \om_3^{i(\l_1+\l_2+\l_3)-1}
\ee

{}For $n=4$, as mentioned before, there is only one energy scale available, therefore in order to study the zero energy limit of one particular particle one needs to treat momentum conservation in a different way than in Refs.\cite{Pasterski:2017ylz,Schreiber:2017jsr,Stieberger:2018edy} and, similarly to $n=3$, relax the kinematics to the $(+-+\, -)$ metric signature. The first step, however, is the same: the four-particle delta function is cast into the form
\begin{align}\label{deltafour}\delta^{(4)}(\omega_1 q_1+\omega_2 q_2&-\omega_3 q_3-\omega_4 q_4)
={4\over \omega_4|z_{14}|^2|z_{23}|^2}\\& \times\, \delta\lf(\omega_1-{z_{24}\zbar_{34}\over z_{12}\zbar_{13}}\omega_4\ri)\ \delta\lf(\omega_2-{z_{14}\zbar_{34}\over z_{12}\zbar_{32}}\omega_4\ri)\ \delta\lf(\omega_3+{z_{24}\zbar_{14}\over z_{23}\zbar_{13}}\omega_4\ri)\,
\delta(r-\bar r)\ ,\nonumber\end{align}
where $r$ is the conformal invariant cross ratio:
\be r={z_{12}z_{34}\over z_{23}z_{41}}\ .\ee
We are interested in the $\lambda_4\to 0$ limit which corresponds to $\om_4 \to 0$. In order to have well-defined expressions we need to make some rearrangements in Eq.(\ref{deltafour}). First,
\be\label{step1}
{1\over |z_{14}|^2 |z_{23}|^2} \d(r-\bar{r})= \d(z_{12} z_{34} \bz_{14} \bz_{23}- \bz_{12} \bz_{34} z_{14} z_{23})\ .
\ee
Next,
\be\label{step2}
\d\Big(\om_3+{z_{24}\bz_{14}\over z_{23}\bz_{13}} \om_4\Big)= \d\Big(\om_4+{z_{23}\bz_{13}\over z_{24}\bz_{14}} \om_3\Big) {z_{23}\bz_{13}\over z_{24}\bz_{14}}
\ee
By taking the soft limit $\om_4\to 0$ of the above delta function, we obtain
\be\label{step3}
{z_{23}\bz_{13}\over z_{24}\bz_{14}} \d \Big({z_{23}\bz_{13}\over z_{24}\bz_{14}} \om_3\Big)= { \bz_{13}\over \om_3 }
 \d(\bz_{13})
\ee
where we chose, as in the $n=3$ case, the delta function support on $z_{ij}\neq 0$ for $i,j=1,2,3$. We also assumed $\omega_3\neq 0$. This step requires  $(+-+\, -)$ signature in order to treat  $z_i,\bz_i$ as independent real variables. In (\ref{step3}) we do not take the limit $\bz_{13}\to0$ in the prefactor because, as we shall see, the numerator will cancel against poles form the rest of the terms.
Proceeding in a similar fashion with the remaining delta functions we obtain
\be \label{step4}
\d\Big(\om_1-{z_{24}\bz_{34}\over z_{12}\bz_{13}} \om_4\Big)= \d\Big(\om_1 - {z_{32}\over z_{12}} \om_3\Big)
\ee
where we used the original delta function of (\ref{step2})  to express $\om_4= - {z_{23}\bz_{13}\over z_{24}\bz_{14}} \om_3$  \footnote{Notice that we cannot take simply $\om_4\to0$ since we have denominator $\bz_{13}$ which goes to zero on the locus of the delta function $\delta(\bz_{13})$. } and ${\bz_{34}\over \bz_{14}}\to 1$ on the locus of (\ref{step3}). Similarly, on the locus of (\ref {step3}), the r.h.s.\ of Eq.(\ref{step1})  becomes
\be\label{step5}
\d(z_{12} z_{34} \bz_{14} \bz_{23}- \bz_{12} \bz_{34} z_{14} z_{23})= {\d (\bz_{12})\over \bz_{14} z_{13} z_{24}}
\ee
After all these cumbersome steps one finds that in the $\omega_4\to 0$ limit, the four-particle delta function (\ref{deltafour}) degenerates to a form similar to  Eq.(\ref{deltathree}). After inserting the four-particle MHV amplitude into the Mellin transform, one ends up with
\ba\label{MHVsoftMellin}
 {\cA}_{--+ (+)}=&& \left( {z_{31} \over z_{34} z_{41}}\right) \left( 4 \prod_{i=1}^3 g(\lambda_i)\right)\lim_{\l_4 \to 0} \left(i \l_4 \int^{\,\prime} d\om_4\  \om_4^{i\l_4-1} \right) \times \\
&&\times\, z_{21}^{1-i(\l_1+\l_2)} z_{23}^{i\l_1-1} z_{31}^{i\l_2-1}\d(\bz_{13})\d(\bz_{23}) \int d\om_3 \ \om_3^{i(\l_1+\l_2+\l_3)-1} \nonumber
\ea
where the prime over the integral indicates that it is restricted to  $\omega_4\approx 0$ region thus excluding the ``ultraviolet'' part that can be eliminated by an explicit cutoff or another regularization. The result is that, as expected,
\be\label{MHVsoftConf}
 {\cA}_{--+ (+)}=(-i)\Big({1\over z_{34}}+ {1\over z_{41}}\Big)   {\cA}_{--+}
\ee

Is it possible to reach the $\lambda_4=0$ limit, Eq.(\ref{MHVsoftConf}) from the four-gluon amplitudes discussed in Refs.\cite{Pasterski:2017ylz,Schreiber:2017jsr,Stieberger:2018edy}?
In the notation of Ref.\cite{Stieberger:2018edy}, the MHV amplitude\footnote{After including proper normalization factors.} is given by
\ba\label{MHVMellin4}
 {\cA}_{--+ +}^{\rm ST}=&& 2\pi \left( 4 \prod_{i=1}^4 g(\lambda_i)\right)\,\delta\big(
 \sum_{j=1}^4\lambda_j\big)\,\delta(r-\bar r)\times\\
&&\times\,
\,\bigg({z_{24}\over \zbar_{13}}\bigg)^{i\lambda_1}\bigg({\zbar_{24}\over z_{13}}\bigg)^{i\lambda_3}\ \bigg({\zbar_{34}\over z_{12}}\bigg)^{i(\lambda_1+\lambda_2)}\bigg({z_{14}\over \zbar_{32}}\bigg)^{i(\lambda_2+\lambda_3)}{r^3\over\zbar_{12}^2\, z_{34}^2}
 \nonumber
\ea
In the limit of $\lambda_4=0$, the $z$-dependent factors are finite, therefore the amplitude is suppressed by the normalization factors and $\cA^{\rm ST}_{--+ (+)}=0$. Does it mean that Eq.(\ref{MHVMellin4}) is wrong? The answer is that in deriving
Eq.(\ref{MHVMellin4}) only one particular, nonsingular solution of kinematic (delta function) constraints was taken into account. The ``boundary'' solutions displayed in Eqs.(\ref{step2})-(\ref{step5}), with $\zbar_{ij}=0$, are not included in Eq.(\ref{MHVMellin4}). We conclude that due to kinematic (over)constraints, the conformal soft limit of four-gluon celestial amplitudes is dominated by such singular contributions. Nevertheless as usual, these contributions originate from the soft energy regions, but are harder to reach than in the case of more particles. They can be interpreted as the contributions of Goldstone modes.

Until this point, our discussion was limited to MHV amplitudes. Since we established the connection between soft conformal and soft limits, we can use soft theorems as in Ref.\cite{Cheung:2016iub}, to show that for any helicity configuration,
\be\label{gensoftConf}
 {\cA}_{J_1,J_2,\dots,J_{n-1} (J_n=+)}=(-i)\Big({1\over z_{(n-1)n}}+ {1\over z_{n1}}\Big)   {\cA}_{J_1,J_2,\dots,J_{n-1}}
\ee

Before turning to Ward identities, let us recall that celestial amplitudes correspond to the correlators of primary conformal fields:
\ba\langle \cO^{a_1}_{\lambda_1 J_1}&&\!\!\!\!\cO^{a_2}_{\lambda_2 J_2}\dots\cO^{a_{n-1}}_{\lambda_{n-1}J_{n-1}}\cO^{a_n}_{\lambda_n J_n}\rangle\ = \nonumber\\[2mm] && =\sum_{\sigma\in S_{n-1}}\cA_{J_1J_2\dots J_{n-1} J_n}^\sigma\,\Tr\big(T^{a_1}
T^{a_{\sigma(2)}}
\dots T^{a_{\sigma(n-1)}}T^{a_{\sigma(n)}}\big)\ ,\label{partiala}
\ea
where $a_i$ are the gauge indices and $T^{a_i}$ are the gauge group generators in the fundamental representation.\footnote{$[T^a,T^b]=i\sum_cf^{abc}T^c$.} The sum extends over all permutations $\sigma$ of $\{2,3,\dots,n\}$ and $\cA^{\sigma}$ are the corresponding partial amplitudes. In the limit of $\lambda=0~(\D=1)$, the primary fields define the following holomorphic and antiholomorphic currents associated to conformally soft gauge bosons:
\be j^a(z)=\cO^a_{0+}(z,\zbar)\ , \qquad \bar j^a(\zbar)=\cO^a_{0-}(z,\zbar)\label{curdef}\ .\ee
in the adjoint representation of the gauge group. We will be discussing Ward identities associated to these currents.

Here, as well in the remainder of the paper, we explicitly considered only the amplitudes associated to the identity permutation because all other partial amplitudes can be discussed in exactly the same way.
Actually, it is easy to see that the soft limit written in Eq.(\ref{gensoftConf}) is valid for all partial amplitudes.
After collecting the soft limits of all partial amplitudes in Eq.(\ref{partiala}), we obtain the Ward identity
\ba\label{Jcorrel}
\langle&&\!\!\!\! j^a(z) \cO^{b_1}_{\lambda_1J_1}(z_1, \bz_1) \cO^{b_2}_{\lambda_2J_2}(z_2, \bz_2) \dots \cO^{b_n}_{\lambda_nJ_n}(z_n, \bz_n) \rangle=~~~~~~~~\\[1mm]&& = \sum_{i=1}^n \sum_{c} \ {f^{ab_ic}\over z-z_i}\langle \cO^{b_1}_{\lambda_1J_1}(z_1, \bz_1) \dots \cO^{c}_{\lambda_iJ_i}(z_i, \bz_i) \dots  \cO^{b_n}_{\lambda_nJ_n}(z_n, \bz_n) \rangle\nonumber
\ea
and a similar identity for the antiholomorphic current. In next section, we discuss these Ward identities in the context of the operator product expansion (OPE).

 \section{OPE for Conformal Primaries and Currents}\label{sec:OPE}
 In this section, we discuss the Ward identity (\ref{Jcorrel}) from a different perspective. In the framework of celestial CFT, the singularities at $z_i=z_j$, when the operators associated to two gauge bosons are inserted at the same points on ${\cal CS}^2$, correspond to the singularities of the operator product expansion (OPE). On the other hand,
at the level of four-dimensional kinematics, $z_i=z_j$ corresponds to $q_i=q_j$, hence to the collinear limit of two momenta, $p_i{\parallel}\,p_j$.

The collinear singularities of Yang-Mills amplitudes are reviewed in section 8 of Ref.\cite{tt}. At the tree level, the leading collinear poles arise from partial amplitudes with adjacent gauge bosons, which we choose to be labelled by $n{-}1,n$. They depend on their respective helicities. For identical helicities,\footnote{The following equations are obtained from the formulas listed in Ref.\cite{tt} by replacing the momentum spinor products $\langle n{-}1\, n\rangle=\sqrt{\omega_{n{-}1}\omega_n} z_{(n{-}1)n}$ and $x=\omega_{n-1}/\omega_P$. Here, we use a notation slightly different from previous sections, by
using superscripts to denote helicity states.}
\ba\cM(1,\dots,n{-}1^+, n^+)\label{col1}& \displaystyle=\frac{1}{z_{(n{-}1)n}}{\omega_P\over\omega_{n{-}1}\omega_n} \cM(1,\dots,P^+)+\dots \\[1mm]\label{col2}
\cM(1,\dots,n{-}1^-, n^-)&\displaystyle =
\frac{1}{\zbar_{(n{-}1)n}}{\omega_P\over\omega_{n{-}1}\omega_n}
 \cM(1,\dots,P^-)+\dots\label{coll}\ea
 The neglected terms are regular in the $z_{(n{-}1)n}=\zbar_{(n{-}1)n}=0$ limit.\footnote{Subleading terms are discussed in Ref.\cite{Stieberger:2015kia}.} In the above equations, $P$ denotes the combined momentum of the collinear pair,\footnote{Without loosing generality, we can assume that both collinear particles are either incoming or outgoing, {\em i.e}. $\epsilon_{n-1}=\epsilon_{n}\equiv\epsilon_P$.}
\be P^\mu= p^\mu_{n{-}1}+p^\mu_n\       =\omega_P\, q_P^\mu\label{colp}\ee
with
\be  \omega_P=\omega_{n{-}1}+\omega_n\ ,\qquad q_P^\mu=q_{n-1}^\mu=q_n^\mu\quad (z_{n-1}=z_n=z_P\ , ~ \zbar_{n-1}=\zbar_n=\zbar_P   )\ .\label{coll}\ee
 For opposite helicities,
\ba \cM(1,\dots,n{-}1^-, n^+)=&&\displaystyle\frac{1}{z_{(n{-}1)n}}{\omega_{n-1}\over\omega_{n}
\omega_P} \cM(1,\dots,P^-)\nonumber \\[1mm] &&\displaystyle +\ \frac{1}{\zbar_{(n{-}1)n}}{\omega_n\over\omega_{n{-}1}\omega_P}
\cM(1,\dots,P^+)+\dots\label{col3}\ea

In order to extract the pole singularities of celestial amplitudes at $z_{n-1}=z_n=z_P\ , ~ \zbar_{n-1}=\zbar_n=\zbar_P$, we insert the collinear limits of  Eqs.(\ref{col1}), (\ref{col2}) and (\ref{col3}) into the Mellin transforms of Eq.(\ref{MellinAmplnw}). At the leading order, it is sufficient to use the momentum-conserving delta functions with the sum of collinear momenta replaced by the combined momentum $P$, as in Eqs.(\ref{colp}) and (\ref{coll}), because the poles are already contained in the invariant matrix elements.

We begin with the case of identical helicities. For the $(++)$ helicity configuration of collinear particles $(n{-1},n)$, we obtain
\ba
 {\cA}_{J_1\dots J_{n-2}(++)}(\lambda_i, z_i, \bz_i)&&\!\!=\left(  \prod_{i=1}^n g(\lambda_i)\right)\left(\prod_{i=1}^{n-2}\int d \omega_i \ \omega_i^{i \l_i} \right)
\int d\omega_{n-1} \omega_{n-1}^{-1+i\lambda_{n-1}}\!\! \int d\omega_{n}
 \omega_{n}^{-1+i\lambda_{n}} \nonumber\\ &&\times\,  \frac{\omega_P}{z_{(n-1)n}}\, \d^{(4)}\big(\sum_{i=1}^{n-2} \e_i \om_i q_i+\epsilon_P\omega_P q_P\big)\,
 \cM(1,\dots, n{-}2, P^+)\label{colamp1}\ea
The integrals over the energies of collinear particles yield
\ba\int\! d\omega_{n-1} \omega_{n-1}^{-1+i\lambda_{n-1}} &&\!\!\!\!\!\!\int \! d\omega_{n}
 \omega_{n}^{-1+i\lambda_{n}}
 \,\omega_P\cdots=\int \!d\omega_{P}\!\int^{\omega_P}_0 \!\!d\omega_{n} (\omega_P-\omega_{n})^{-1+i\lambda_{n-1}}\omega_{n}^{-1+i\lambda_{n}}
 \omega_P\cdots\nonumber\\[1mm] &&=~B(i\lambda_{n-1},i\lambda_n)\int \!d\omega_{P}\,\omega_P^{i\lambda_{P}}\cdots\ ,\label{intpp}
\ea
where
\be\lambda_P=\lambda_{n-1}{+}\lambda_n\ee
and $B$ denotes the Euler's beta function,
\be B(x,y)=\frac{\Gamma(x)\Gamma(y)}{\Gamma(x+y)}.\ee
As a result, we obtain
\ba &&
{\cA}_{J_1\dots J_{n-2}(++)}(\lambda_i, z_i, \bz_i)=(-i)
\frac{C_{(++)+}(\lambda_{n-1},\lambda_n)}{z_{n-1}-z_n}~~~~~~\nonumber
\\[2mm] && ~~~\times\,
{\cA}_{J_1\dots J_{n-2}+}(\lambda_1,\dots, \lambda_{n-2},\lambda_P;
z_1,\dots,z_{n-2},z_P;\zbar_1,\dots,\zbar_{n-2},\zbar_P   )\ea
where, with the normalization factors given in (\ref{gdef}),
\be C_{(++)+}(\lambda_{n-1},\lambda_n)
~=~i\,\frac{g(\lambda_{n-1})g(\lambda_{n})}{g(\lambda_{n-1}+\lambda_n)}
B(i\lambda_{n-1},i\lambda_n)~=~1+\frac{\lambda_{n-1}\lambda_{n}}{(1+
i\lambda_{n-1})(1+i\lambda_{n})}\ .\ee

After collecting the collinear poles of all partial amplitudes in Eq.(\ref{partiala}),
we find the  leading OPE terms:
\be\label{opepp}
\cO^{a}_{\lambda_1 +}(z,\zbar)\,\cO^{b}_{\lambda_2+}(w,\bar w)
=\frac{C_{(++)+}(\lambda_1,\lambda_2)}{z-w}\sum_c f^{abc}\cO^{c}_{(\lambda_1+\lambda_2)+}(w,\bar w)+\dots,
\ee
where
\be C_{(++)+}(\lambda_1,\lambda_2)=1+\frac{\lambda_1\lambda_2}{(1+i\lambda_1)(
1+i\lambda_2)}\label{ope1}\ee
The case of identical $(--)$ helicity configurations can be discussed in the same way, leading to
\be\label{opemm}
\cO^{a}_{\lambda_1 -}(z,\zbar)\,\cO^{b}_{\lambda_2-}(w,\bar w)
=\frac{C_{(--)-}(\lambda_1,\lambda_2)}{\zbar-\bar w}\sum_c f^{abc}\cO^{c}_{(\lambda_1+\lambda_2)-}(w,\bar w)\dots,
\ee
with
\be C_{(--)-}(\lambda_1,\lambda_2)= C_{(++)+}(\lambda_1,\lambda_2)
\label{ope2}\ee

For opposite helicities, the collinear limits of Eq.(\ref{col3}) lead to energy integrals different from (\ref{intpp}). Instead, one finds
\ba\int\! d\omega_{n-1} \omega_{n-1}^{1+i\lambda_{n-1}} &&\!\!\!\!\!\!\int \! d\omega_{n}
 \omega_{n}^{-1+i\lambda_{n}}
 \frac{1}{\omega_P}\cdots=\int \!d\omega_{P}\!\int^{\omega_P}_0 \!\!d\omega_{n} (\omega_P-\omega_{n})^{1+i\lambda_{n-1}}\omega_{n}^{-1+i\lambda_{n}}
 \frac{1}{\omega_P}\cdots\nonumber\\[1mm] &&=~B(2+i\lambda_{n-1},i\lambda_n)\int \!d\omega_{P}\,\omega_P^{i\lambda_{P}}\cdots\ .\label{intpm}
\ea
In this case,
\be i\, \frac{g(\lambda_{n-1})g(\lambda_{n})}{g(\lambda_{n-1}+\lambda_n)}
B(2+i\lambda_{n-1},i\lambda_n)~=~\frac{\lambda_{n-1}}{(\lambda_{n-1}
+\lambda_{n})(1+i\lambda_n)}\ .\ee
After repeating the same steps as for identical helicities, we find
the following leading OPE terms
\ba
\cO^{a}_{\lambda_1 -}(z,\zbar)\,\cO^{b}_{\lambda_2+}(w,\bar w)
~=~&&\!\!\!\frac{C_{(-+)-}(\lambda_1,\lambda_2)}{z-w}\sum_c f^{abc}\cO^{c}_{(\lambda_1+\lambda_2)-}(w,\bar w)~~~~\nonumber\\[2mm]
&&+\,\frac{C_{(-+)+}(\lambda_1,\lambda_2)}{\zbar-\bar w}\sum_c f^{abc}\cO^{c}_{(\lambda_1+\lambda_2)+}(w,\bar w)+\dots\label{opemp}
\ea
with
\ba C_{(-+)-}(\lambda_1,\lambda_2)&=&\frac{\lambda_1}{(\lambda_1
+\lambda_2)(1+i\lambda_2)}\label{ope3}\\[1mm]
C_{(-+)+}(\lambda_1,\lambda_2)&=&\frac{\lambda_2}{(\lambda_1
+\lambda_2)(1+i\lambda_1)}\label{ope4}
\ea

The OPE for the holomorphic current defined in Eq.(\ref{curdef}) follows from Eqs.(\ref{opepp}) and (\ref{opemp}) by setting $\lambda_2=0$ in the coefficients (\ref{ope1}), (\ref{ope3}) and (\ref{ope4}). The result is
\be\label{opejh}
j^{a}(z)\,\cO^{b}_{\lambda J}(w,\bar w)
=\frac{1}{z-w}\sum_c f^{abc}\cO^{c}_{\lambda J}(w,\bar w)+\dots
\ee
Similarly, by seeting $\lambda_1=0$ in the coefficients (\ref{ope2}), (\ref{ope3}) and (\ref{ope4}), we obtain
\be\label{opeja}
\bar j^{a}(\zbar)\,\cO^{b}_{\lambda J}(w,\bar w)
=\frac{1}{\zbar-\bar w}\sum_c f^{abc}\cO^{c}_{\lambda J}(w,\bar w)+\dots
\ee
These OPE rules lead to the same Ward identity (\ref{Jcorrel}) as the soft limits considered in the previous section. Acting on the vacuum, the currents (\ref{curdef}) create Goldstone modes. Since
\be\label{opejj}
j^{a}(z)\,j^b(w)
=\frac{1}{z-w}\sum_c f^{abc}j^c(w)+\dots
\ee
these currents are associated to the nonabelian (global) symmetry of CFT on celestial sphere.

Finally, let us recall that throughout all computations, we used the $(-i)^\D$ prefactor in the normalization factors $g(\D)$, see the remark after Eq.(\ref{gdef}). This means that the OPE rules written above apply to operators creating {\em outgoing} particles. For incoming particles, there is a minus sign appearing on the r.h.s.\ of Eqs.({\ref{opepp}), ({\ref{opemm}) and ({\ref{opemp}). After conjugating these products, we find that
\be \cO^{a, \rm in}_{\lambda\, J}(z,\zbar)=\big[\cO^{a, \rm out}_{{-}\!\lambda\, {-}\!J}(z,\zbar)\big]^\dagger\ .\ee
In this way, four-dimensional CPT symmetry is implemented at the level of celestial CFT.

\section{Conclusions}\label{conc}
In this paper, we used traditional field-theoretical techniques to show that in the conformally soft limit ($\lambda=0$) of primary fields associated to gauge bosons, celestial amplitudes are dominated by the contributions of soft ($\omega\to 0$) particles, thus confirming the same conclusion reached in the ``holographic'' framework of quantum field theory in curvilinear coordinates. We also derived the OPE rules for these primary field operators and the currents that appear in their conformally soft limit and showed that the Ward identities associated to the currents have the same form as the identities obtained by using soft theorems. As a corollary, Yang-Mills soft theorems follow from collinear theorems  in the same way as Ward identities follow from OPE in two-dimensional CFT.

Celestial CFT of primary fields in the principal series with complex dimensions seems rather exotic, but after all it is yet another realization of four dimensional Yang-Mills theory. At the tree level, Yang-Mills is a simple theory, so it is a reasonable expectation that at least some aspects of celestial CFT are not too complicated. The OPE derived in this work should provide a good starting point for further investigations of celestial CFT.

\vskip 1mm
\leftline{\noindent{\bf Acknowledgments}}
\vskip 1mm
\noindent
We are grateful to Monica Pate, Ana Raclariu and Andy Strominger for sharing a draft of Ref.\cite{prs} before its publication. We also thank them and Ellis Yuan for very useful conversations and suggestions.
This material is based in part upon work supported by the National Science Foundation
under Grant Number PHY--1620575.
Any opinions, findings, and conclusions or recommendations
expressed in this material are those of the authors and do not necessarily
reflect the views of the National Science Foundation.

\renewcommand{\thesection}{A}
\setcounter{equation}{0}
\renewcommand{\theequation}{A.\arabic{equation}}
\renewcommand{\thesection}{A}
\setcounter{equation}{0}
\renewcommand{\theequation}{A.\arabic{equation}}
\vskip 3mm
\begin{center}{\noindent{\large\bf Appendix}}\end{center}\noindent
Our goal is to solve the $n$-particle momentum conservation constraints ($n\ge 5)$ in a way that allows straightforward access to $n{-}5$ soft regions of zero energy. To that end, we define the following cross-ratios:
\begin{equation}
t_i = \frac{z_{12}z_{3i}}{z_{13}z_{2i}}, \quad i=4,5,\ldots, n.
\end{equation}
The $n$-point momentum delta function can be written as
\begin{equation}
\delta^4\big(\sum_{i=1}^{N} \epsilon_i \omega_i q_i\big) = \frac{i}{4}\frac{(1-t_4)(1-\bar{t}_4)}{t_4-\bar{t}_4} \frac{1}{|z_{14}|^2|z_{23}|^2} \prod_{i=1}^{4}\delta(\omega_i-\omega_i^\star).
\end{equation}
The solutions are
 \be \omega_i^\star=f_{i5}\omega_5+f_{i6}\omega_6+\ldots+f_{in}\omega_n\ ,\ee where $f_{ij}, i=1,2,3,4, j=5,6,\ldots,n$ are the following  functions of celestial coordinates:
\begin{align}
	f_{1j}&=t_4 \Big|\frac{z_{24}}{z_{12}}\Big|^2 \frac{(1-t_4)(1-\bar{t}_4)}{t_4-\bar{t}_4}\epsilon_1\epsilon_j\frac{t_j-
\bar{t}_j}{(1-t_j)(1-\bar{t}_j)}\Big|\frac{z_{1j}}{z_{14}}\Big|^2   - \epsilon_1\epsilon_j t_j\Big|\frac{z_{2j}}{z_{12}}\Big|^2 \\
	f_{2j}&= - \frac{1-t_4}{t_4} \Big|\frac{z_{34}}{z_{23}}\Big|^2 \frac{(1-t_4)(1-\bar{t}_4)}{t_4-\bar{t}_4}\frac{\epsilon_1\epsilon_j}{\epsilon_1
\epsilon_2}\frac{t_j-\bar{t}_j}{(1-t_j)(1-\bar{t}_j)}\Big|\frac{z_{1j}}{z_{14}}\Big|^2 + \frac{\epsilon_1\epsilon_j}{\epsilon_1\epsilon_2} \frac{1-t_j}{t_j}\Big|\frac{z_{3j}}{z_{23}}\Big|^2\\
	f_{3j}&= (1-t_4)\Big|\frac{z_{24}}{z_{23}}\Big|^2 \frac{(1-t_4)(1-\bar{t}_4)}{t_4-\bar{t}_4}\frac{\epsilon_1\epsilon_j}{\epsilon_1
\epsilon_3}\frac{t_j-\bar{t}_j}{(1-t_j)(1-\bar{t}_j)}\Big|\frac{z_{1j}}{z_{14}}\Big|^2 - \frac{\epsilon_1\epsilon_j}{\epsilon_1\epsilon_3} (1-t_j)\Big|\frac{z_{2j}}{z_{23}}\Big|^2 \\
	f_{4j}&= - \frac{(1-t_4)(1-\bar{t}_4)}{t_4-\bar{t}_4}\frac{\epsilon_1\epsilon_j}{\epsilon_1
\epsilon_4}\frac{t_j-\bar{t}_j}{(1-t_j)(1-\bar{t}_j)}\Big|\frac{z_{1j}}{z_{14}}\Big|^2
\end{align}

\end{document}